\documentclass[preprint,authoryear]{elsarticle}
\usepackage{import}
\usepackage[top=2cm,bottom=2cm,left=2cm,right=2cm]{geometry} 
\usepackage[utf8x]{inputenc}
\usepackage{amsmath, amssymb, amsfonts, amsthm, mathtools}
\usepackage{mathrsfs}  

\usepackage{multicol}
\usepackage{ulem}

\usepackage[title]{appendix}

\usepackage{float}
\usepackage{tikz}
\usetikzlibrary{calc}

\usepackage{enumitem}
\usepackage{hyperref}
\everymath{\displaystyle}

\usepackage{multicol} 
\usepackage{multirow}

\newdefinition{definition}{Definition}
\newdefinition{theorem}{Theorem}
\newdefinition{claim}{Claim}
\newdefinition{lemma}{Lemma}
\newdefinition{step}{Step}[theorem]
\newdefinition{remark}{Remark}
\newdefinition{proposition}{Proposition}
\newdefinition{corollary}{Corollary}[theorem]
\newdefinition{example}{Example}
\newdefinition{assumption}{Assumption}
\newdefinition{condition}{Condition}

\usepackage{setspace}
\usepackage{array}
\newcommand{\PreserveBackslash}[1]{\let\temp=\\#1\let\\=\temp}
\newcolumntype{C}[1]{>{\PreserveBackslash\centering}p{#1}}
\newcolumntype{R}[1]{>{\PreserveBackslash\raggedleft}p{#1}}
\newcolumntype{L}[1]{>{\PreserveBackslash\raggedright}p{#1}}

\begin{document}

\journal{Arxiv.org}

\begin{frontmatter}
	\title{Equilibria of Attacker-Defender Games}
	
	\author[1]{Zsombor Z.~M\'eder\corref{cor1}}
	\ead{z.z.meder@fsw.leidenuniv.nl}
	\author[1]{Carsten K.W. de Dreu}
	\ead{c.k.w.de.dreu@fsw.leidenuniv.nl}
	\author[2]{J\"org Gross}
	\ead{mail@joerg-gross.net}
	
	\address[1]{Institute of Psychology, Leiden University, Netherlands.}
	\address[2]{Department of Psychology, University of Zürich, Switzerland.}
	\cortext[cor1]{Corresponding author}

	\begin{abstract}
		\baselineskip=1.3\baselineskip
		\noindent
		
		Attempts at predatory capture may provoke a defensive response that reduces the very value of the predated resource. We provide a game-theoretic analysis of simultaneous-move, two-player Attacker-Defender games that model such interactions. When initial endowments are equal, Attackers win about a third of such games in equilibrium. Under power disparities, Attackers become particularly aggressive when they are approximately one-third poorer than Defenders. With non-conflictual outside options Attackers become exceptionally aggressive when their opponent has access to a high-benefit, low-cost production option, and refrain from attack most when they are unilaterally provided with a high-benefit, high-cost production option. 
		
	\end{abstract}
\end{frontmatter} 

\section{Introduction}

Situations of conflict involve parties with antagonistic interests. One way of modeling such conflicts is to describe them as contests, with parties expending effort in order to win a reward. For example, in a Tullock contest, players may gain a fixed prize, and their probability  to win depends on the relative effort that they exert. In many setups, these efforts are irretrievable, regardless of whether a player wins or loses. The literature on all-pay auctions focuses on such contexts, and has numerous applications in modelling R$\&$D races, labor-market competition, political lobbying, and law. See \cite{siegel09, siegel10} for a thorough analysis of a large class of such games.

In simple all-pay auctions, the winner's payoff depends only on their effort and their valuation of the prize, which is given ex ante. However, in many real-life applications, players' efforts invested in the contest may influence the value of the prize in itself. For example, in sporting competitions, tighter contests may lead to a more exciting competition, larger viewership, and consequently better future sponsorship deals for the winner. In such cases, the effort invested creates positive spillovers. Such setups have been used, for instance, for modelling R$\&$D races \citep{che03, zhou06} and labour market competition \citep{lazear81}. Sometimes, however, effort invested creates negative spillovers. A classic example is provided by wars of attrition \citep{smith74}, in which the loser's effort decreases the magnitude of the prize.

The literature on such phenomena, in which the contestant's efforts has spillovers, has been recently expanding \citep{baye12, xiao18, betto21}. This literature speaks to, but does not yet fully cover conflicts that arise because one side seeks to acquire resources already held and protected by another party \citep{dreu19, durham76, miller09, blattman10}. From a general perspective, these are conflicts in which the pre-contest status quo in terms of who controls the prize is important: one party aims to maintain it, while the other attempts to overturn it. Examples include countries resorting to scorched-earth tactics when under attack, destroying the spoils of war sought after by their aggressors \citep{lacomba17}. Another example concerns owners using company reserves to hold out against union strikers, reducing what is available for wage adjustment demanded by employees. Finally, predator-prey interactions and conspecifics attempting to capture territory or mates can be understood as situations in which the protective efforts of a defender can reduce what an attacker can obtain \citep{dreu22, abrams00, arnott09}.

How players in such asymmetric contests spend effort at predation and concomitant defensive responses are poorly understood. Here we develop the equilibrium properties of such games of `attack' and `defense', providing a framework for understanding and studying the mechanisms underlying such conflicts. Similar to all-pay contests, our model involves non-retrievable costs. In contrast to these, however, we assign two different roles (`Attacker' and `Defender') to the involved parties, capturing the asymmetry of the ex ante situation. Attempts at maintaining the status quo reduce the value of the contested resource for both sides. Our setting thus involves negative (direct) spillovers \citep[see][]{baye12} for Attacker in case they win. For Defender, losing reduces their payoff to zero, their payoff structure is thus some-or-nothing. This corresponds to the often dramatic consequences of losing a war, ownership of a firm, or -- in the case of biological prey -- the life of the organism.

As noted above, our analysis is related to an extensive literature on contests and all-pay auctions \citep[for overviews, see][]{long13, dechenaux15, siegel09, siegel10}, and more specifically, to the work on all-pay auctions with spillovers \citep[]{baye12, xiao18, betto21}.  Our model is distinguished by three features. First, due to interpreting the status quo as significant, we impose an asymmetry in the payoff structure between Attacker and Defender. The incentive for conflict on Attacker's side arises from the value of the resources held by Defender; Defender's goal, in contrast, is to repel the predatory attack and maintain ownership of these resources. Second, as outlined above, Defender's effort imposes a negative spillover in case Attacker wins, but not vice-versa. Third, we work in a discrete setting, for two reasons: it matches how resources are quantized in the real world -- a gang cannot, for example, commit $3.4$ members to protect its territory, nor can an aggressor engage $17.8$ tanks in an attack. This has the additional advantage to provide a more direct and practical benchmark for experiments that use an Attacker-Defender setup.

Our paper is structured as follows. In Section \ref{section:equal}, we describe the basic structure and  equilibria of Attacker-Defender conflicts, and reveal that Attackers will succeed in their predatory efforts in slightly more than one out of three interactions. However, and illustrating how surprising such scenarios are from a game theoretic perspective, Attackers refrain from predatory efforts altogether in a similar share of cases.

Sections \ref{section:unequal} and \ref{section:outside} examine specific cases of Attacker-Defender conflict. In Section \ref{section:unequal} we investigate the mixed-strategy equilibrium when contestants differ ex ante in wealth and concomitant fighting capacity. We find that conflict can be more intense when Attacker is less wealthy than Defender, resonating with work on the `paradox of power' \citep{hirshleifer91, durham98}. This analysis also reveals how `leveling the playing field' among contestants by third parties could provide an effective measure to reduce conflict. We further show that similar dynamics emerge when introducing psychological motives like a `warm-glow' from winning a contest, as discussed in the empirical conflict or auction literature \citep[e.g.,][]{bos08}.

Section \ref{section:outside} investigates how the mixed-strategy equilibrium changes when contestants have outside options for the production of economic wealth (i.e., possibilities to obtain a prize through non-conflictual means like technological innovation, extraction of natural resources, or agricultural/industrial production). Here, our analysis integrates insights from work suggesting that opportunities for economic production should lower the incentives to engage in conflict \citep{maoz93, rousseau96, wittman00} with studies on the natural resource curse and the `guns versus butter' dilemma that suggests that wealth disparities can trigger aggressive appropriation and investment in protective defense  \citep{ploeg11, brunnschweiler09, grossman96, carter01}. We specify the conditions under which outside options increase or decrease conflict expenditures and, as a result, increase or decrease post-conflict wealth disparities among contestants. Our analysis shows, among other things, that players with relatively inexpensive production facilities invest more in predatory capture. Players with relatively expensive production options, in contrast, substantially reduce efforts at predation. The reason behind this is that predation and production become substitutes when they cannot be successfully pursued simultaneously. Hence, these results show that providing potential Attackers with high-cost, high-yield investment opportunities can contribute to the reduction of social waste generated by conflict.

Section \ref{section:conclusion} concludes with limitations and with possible avenues for future research and theory development. 

\section{Attack and defense under equal endowments}\label{section:equal}

Our analysis concerns asymmetric conflict between two players, who both start with some initial endowment. The two sides' strategic decisions concern the part of the endowment to be spent on conflict, since the conflict will be won by the side making a larger investment. The interaction is one-off, and decisions are made simultaneously. This defines the Attacker-Defender (A-D) game as a one-shot, perfect and complete information game of two players: $\Gamma_{AD}=\left\{\{A,D\},(S_A,S_D),(\pi_A,\pi_D)\right.\}$. We specify the strategy sets and payoff functions in each following section. 

\subsection{Equilibrium}\label{subsection:equilibrium}

Assume first that both sides have a starting endowment of $R\geq 2$, and that they can invest any integer value up to $R$. The two players' strategy sets are $S_A=S_D=\{0,1,\dots,R\}$, while the payoff functions are given by:
\[\pi_A(i,j)=
\begin{cases}
	2R-i-j & i>j,\\
	R-i& i\leq j;
\end{cases}\text{~~~~~and~~~~~}
\pi_D(i,j)=
\begin{cases}
	0 & i>j,\\
	R-j& i\leq j,
\end{cases} \text{~with~} i\in S_A,~j\in S_D.\]

Table \ref{fig:general} represents such games in matrix form. A brief consideration of the matrix and the payoffs is sufficient to establish that these games possess no pure-strategy equilibria. Defender's best-response is to match Attacker's investment. When investments are equal, however, Attacker's best response is either to completely abstain from attacking, or to invest exactly one unit more than Defender, depending on the magnitude of the potential spoils (see Lemma \ref{lemma:only-mixed} in the Appendix).

\begin{table}[!ht]
	\centering
	\begin{tabular}{rc|C{2cm}|C{2cm}|C{2cm}|c|C{3.2cm}|c|C{2cm}|}
		&\multicolumn{1}{c}{} &\multicolumn{7}{c}{Defender}\\
		&\multicolumn{1}{c}{} & \multicolumn{1}{c}{$0$} & \multicolumn{1}{c}{$1$} & \multicolumn{1}{c}{$2$} & \multicolumn{1}{c}{\dots} & \multicolumn{1}{c}{$j$}& \multicolumn{1}{c}{\dots} & \multicolumn{1}{c}{$R$} \\ \cline{3-9}
		\multirow{7}{*}{Attacker} 
		& $0$    & $R,R$ & $R,R-1$ & $R,R-2$ & & $R,R-j$ & & $R,0$\\ \cline{3-9}
		& $1$    & $2R-1,0$ & $R-1,R-1$ & $R-1,R-2$ & & $R-1,R-j$ & & $R-1,0$\\ \cline{3-9}		
		& $2$    & $2R-2,0$ & $2R-3,0$ & $R-2,R-2$ & & $R-2,R-j$ & & $R-2,0$ \\ \cline{3-9}			
		& $\dots$  & & & & & & & \\ \cline{3-9}
		& $i$    & $2R-i,0$ & $2R-i-1,0$ & $2R-i-2,0$ & & $
		\begin{array}{lr}
			2R-i-j,0 & : i>j\\
			R-i,R-j& : i\leq j 
		\end{array}$ & & $R-i,0$ \\ \cline{3-9}
		& $\dots$  & & & & & & & \\ \cline{3-9}
		& $R$ & $R,0$  & $R-1,0$ & $R-2,0$ & & $R-j,0$ & & $0,0$ \\ \cline{3-9}
	\end{tabular}
	\caption{A-D games with symmetric resources and discrete unit control in matrix form.}\label{fig:general}
\end{table}	

Mixed strategies will be indicated as $a=(a_i)_{i\in \{0,\dots,R\}}$, and  $d=(d_j)_{j\in \{0,\dots,R\}}$. The function $supp(s)$ gives the set of indices in the support of a mixed strategy. Expected conflict investments for a strategy $s$ will be denoted by $E(s)$, while expected payoffs by $\Pi_A$ and  $\Pi_D$. We use $W$ to indicate the event that Attacker wins, and $P(W)$ for its probability. Proofs are provided in the Appendix.

\begin{theorem}\label{theorem:equilibrium}
	The A-D game with equal endowments of $R$ has a single, mixed-strategy Nash equilibrium $(a^*,d^*)$, with:
	\[a^*_i=\begin{cases}
		\displaystyle \frac{R-l^*}{R}&i=0,\\
		\displaystyle \frac{R-l^*}{(R-i+1)(R-i)}&1\leq i\leq l^*,\\
		0&l^*<i\leq R;
	\end{cases}\text{~~~~~and~~~~~}
	d^*_j=\begin{cases}
		\displaystyle \frac{1}{R-j}& j<l^*,\\
		\displaystyle 1-\sum_{i=0}^{l^*-1}\frac{1}{R-i}& j=l^*,\\
		0& l^*<j \leq R.
	\end{cases}\]
	
	\noindent The value of $l^*$ is identified by: \[H_{R}-H_{R-l^*} < 1 < H_{R}-H_{R-l^*-1},\] \noindent where 
	$H_i$ denotes the harmonic series up to $i$: $H_i=_{\text{def}}1+\frac{1}{2}+\frac{1}{3}+\dots+\frac{1}{i}.$
	
	 Thus, $l^*$ is the boundary where the series $\displaystyle \frac{1}{R}+\frac{1}{R-1}+\dots$ would exceed $1$. We call $l^*$ the `upper bound' (of equilibrium conflict investments).
	
\end{theorem}

\begin{corollary}\label{total-earnings}
	The sum of expected payoffs is equal to the total of endowments less the upper bound:
	\[\Pi_A+\Pi_D=R-E(a^*)+R-E(d^*)=2R - l^*.\]	
\end{corollary}

\begin{corollary}\label{conflict-investments}
	The sum of expected conflict investments is equal to the upper bound: \[E(a^*)+E(d^*)=l^*.\]
\end{corollary}

\begin{theorem}\label{upper-bound-limit} 
	The ratio of the upper bound to endowment converges to  $\displaystyle 1-\frac{1}{e}$ as the endowment goes to infinity:
	\[\displaystyle \lim_{R\to \infty} \frac{l^{*}}{R}=1-\frac{1}{e}.\]
\end{theorem}

\begin{corollary}\label{earnings-ratio-limit}
	Attacker's share of earnings converges to $\displaystyle \frac{e}{e+1}$:
	\[\displaystyle \lim_{R\to \infty} \frac{\Pi_A}{\Pi_A+\Pi_D}=\frac{e}{e+1}.\]
\end{corollary}

\begin{corollary}\label{conflict-investments-limit}
	Expected conflict investment proportional to endowment converge to  $\displaystyle 1-\frac{2}{e}$ for Attacker, and $\displaystyle \frac{1}{e}$ for Defender as the endowment goes to infinity:
	\[\lim_{R\to \infty} \frac{E(a^*)}{R}=1-\frac{2}{e}\text{~~~~~and~~~~~} \lim_{R\to \infty} \frac{E(d^*)}{R}=\frac{1}{e}.\]
\end{corollary}

\begin{corollary} Defender's expected conflict investment proportional to their endowment when Attacker wins (loses) converges to $3-e$ $\left(\text{respectively, } \displaystyle1-\frac{1}{e-1}\right)$:
	\[\displaystyle \lim_{R\to \infty} \frac{E(d|W)}{R}=3-e \text{~~~~~and~~~~~}
	\lim_{R\to \infty} \frac{E(d|L)}{R}=1-\frac{1}{e-1}.\]
\end{corollary}

\begin{theorem}\label{winrate-limit}
	The probability of Attacker winning converges to $\displaystyle \frac{1}{e}$ as the amount of endowment goes to infinity:
	\[\displaystyle \lim_{R\to \infty} P(W)=\frac{1}{e}.\]
\end{theorem}

\subsection{Interpretation}\label{eq:interpretation}

Consider the circularity of best-responses in the A-D game with equal endowments of $R$. If Attacker is winning the conflict, Defender needs to match Attacker's conflict investment; but then Attacker is better off by abstaining from conflict; which would allow Defender to lower its guard; which makes attacking worthwhile. The lack of a pure-strategy equilibrium means that the emergence and intensity of attack and defense cannot be precisely predicted by the opponent or outside observers.

While the previous observation is already sufficient to show the non-existence of a pure-strategy equilibrium, it does not yet reveal the precise structure of the mixed equilibrium. To obtain this, we note that the support of equilibrium strategies includes all integer investments from $0$ up to a certain upper bound $l^*$, the highest conflict investment used by both Attacker and Defender. The easiest way to determine the upper bound is from the equilibrium mix of Defender. This assigns weights $\frac{1}{R}, \frac{1}{R-1}, \dots, \frac{1}{R-l^*}$  to choosing $0, 1, \dots, l^*$. Once Defender's `probability allotment' of $1$ is about to be exhausted, the remaining probability weight is put on strategy $l^*$. The upper limit is thus determined by the point where the `reverted harmonic sum' starting at $\frac{1}{R}$ exceeds $1$. Theorem \ref{upper-bound-limit} shows that in the limit, this is at $1-\frac{1}{e}\approx 63\% $ of the endowment. Neither predation, nor defense costing more than two-thirds of the endowment will be used in equilibrium.

\begin{figure}[ht!]
	\center
	\includegraphics{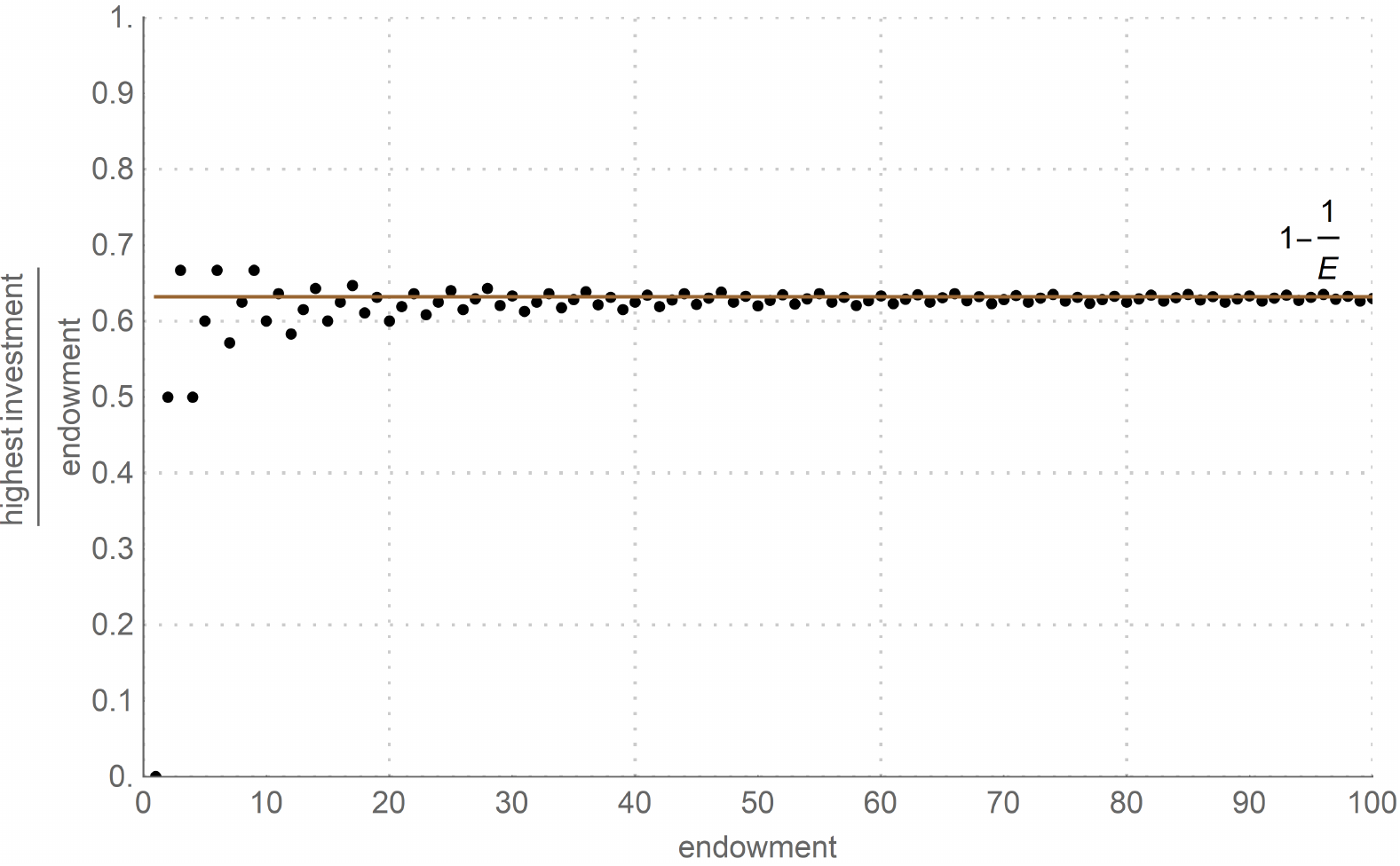}
	\caption{The upper limit $l^*$, corresponding to the highest attack/defense investment as a share of the initial endowment, showing convergence to $1-1/E$.}\label{fig:equal-lstar}
\end{figure}

Next, we see that \textit{all} actions up to the upper bound will be used with some probability. This implies, inter alia, that both players have a positive probability of choosing $0$. Thus, conflict may be entirely absent, and the outcome that maximizes social efficiency -- interpreted here as the sum of payoffs -- occurs with positive probability. At the same time, the probability for Attacker to choose fully abstaining from conflict (i.e., choosing $0$) will be higher than for Defender. The former approaches $\frac{1}{e}$ in the limit, while the latter vanishes. This corresponds to a difference between the two sides' mixes: Defender assigns higher probabilities to larger conflict investments. Such monotonicity only holds for choosing actions $1, 2, ..., l^*$ for Attacker; for them, the most likely action is to completely refrain from attacking.

Average conflict investments are higher for almost all values of $R$. In the limit, $\displaystyle \frac{1-\frac{1}{e}}{2}\approx 32\%$ of all initial endowments are spent on conflict, as Defender invests about $38\%$ more than Attacker, or, equivalently, $10$ percentage points more of their endowment (Defender vs. Attacker: $\approx36\%$ vs. $\approx27\%$ of $R$).

\begin{figure}[ht!]
\center
\includegraphics{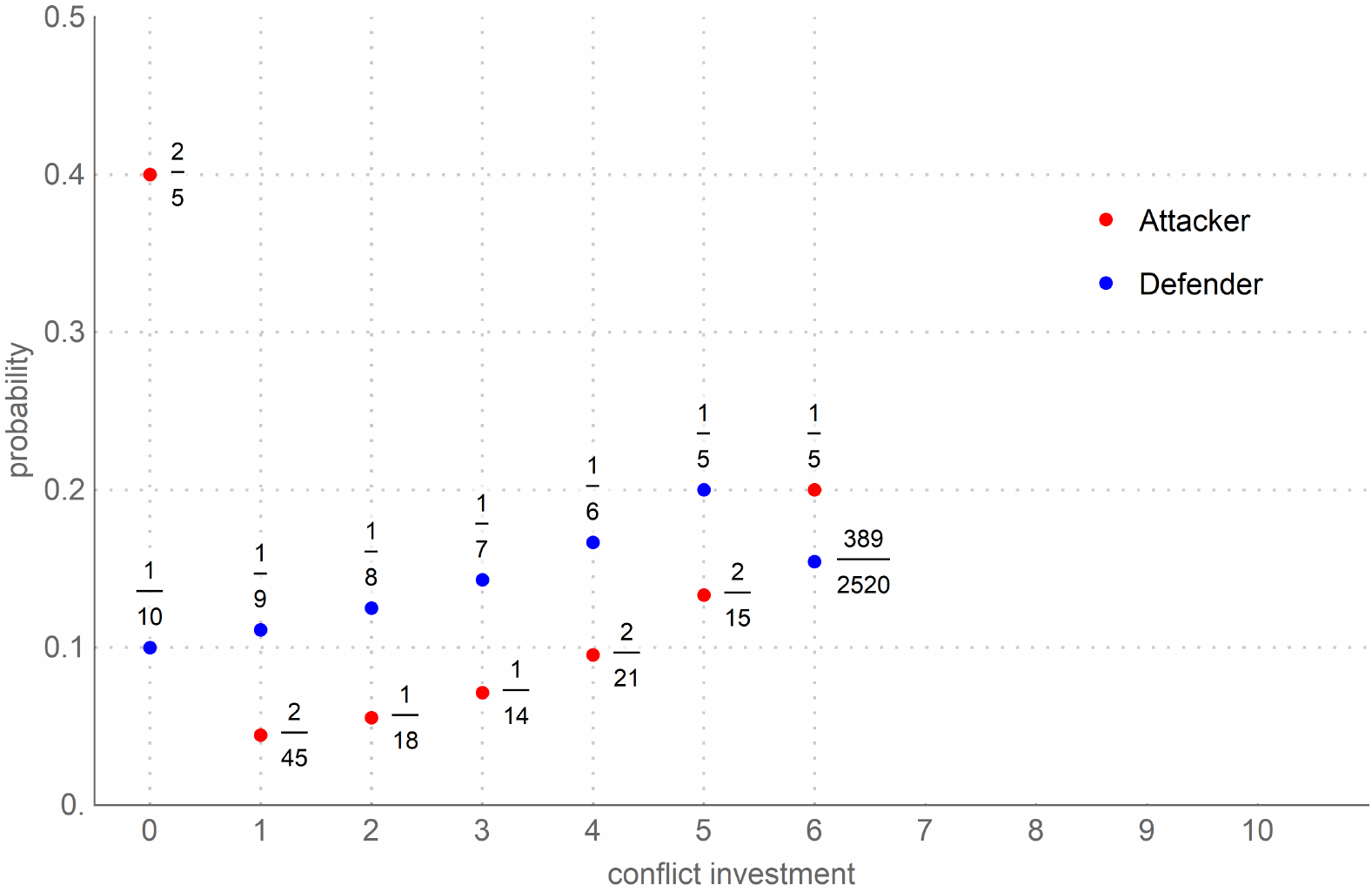}
\caption{Equilibrium strategies for equal endowments of $R=10$.}\label{fig:r10}
\end{figure}

Theorem \ref{winrate-limit} indicates that Attacker wins, in the limit, about $\displaystyle \frac{1}{e}\approx 37\%$ of interactions. Combining this result with the frequency of abstentions from attack yields an interesting typology of Attacker-Defender interactions: With a probability of $\approx37\%$, Attacker is passive; in a further $37\%$ of interactions, Attacker successfully wins the contest; while only in the remaining $26\%$ of all cases does Defender successfully defend against an actual attack. In other words, Defender's apparently high success rate emerges because the equilibrium mix for Attacker often prescribes indolence.  It is interesting to note the analogy with the life-dinner principle identified in evolutionary biology, suggesting that (a) predation is less forceful overall than prey-defence, and (b) predatory success is infrequent enough for prey to adapt and reproduce \citep[see][]{dawkins79, vermeij82}. 

Since the expected payoffs of all strategies in the support of the mix need to be equal, earnings on both sides can be determined by the value of the respective `safe options'. For Attacker, this is refraining from attack, yielding $R$; while for Defender, this is defending at the upper bound, yielding $R-l^*$. Corollary \ref{earnings-ratio-limit} shows that in the limit, Attacker will, in expectation, earn about close to three quarters of the endowment remaining after conflict. Total earnings, according to Corollary \ref{total-earnings}, are equal to $2R-l^*$. The upper bound $l^*$ turns out to correspond not only to the highest conflict investment used by the two sides, but also to the expected share of endowments consumed by conflict. As a proportion of initial endowments, $\displaystyle \frac{l^*}{2R}$ resources are wasted in expectation, i.e., $\approx 32\%$. This quantifies the observation by \cite{mill48} that ``disproportionate amount of energies are spent on injuring others and protecting against injury.'' Indeed, in equilibrium, this `disportionate' amount is about one-third of the players' total wealth.

\subsection{Tie-breaking rules}\label{subsection:tiebreaking}

Recall that in the model presented in Section \ref{section:equal}, Attacker loses and thus, Defender keeps their remaining endowment in case of equal conflict contributions. A straightforward alternative to this conflict success function is to flip this rule, so that Attacker is victorious in case of a tie. Generally, Attacker might win a tie with a fixed probability $t\in [0,1]$. Using this notation, the model in Section \ref{subsection:equilibrium} assumes $t=0$. 

To understand how changing the tie-breaking rule alters the equilibrium, let us switch $t=0$ into $t=1$ first. This turns $d=0$ into a dominated strategy for Defender. In effect, $d=1$ would perform the same function under $t=1$ as $d=0$ under $t=0$. Generally, choosing $d=i+1$ protects Defender successfully under $t=1$ against exactly those attacks as $d=i$ does under $t=0$, at the cost of $1$ unit of payoff. By switching from $t=0$ to $t=1$, equilibrium mixes merely `shift' by one unit:

\[a^*_i=\begin{cases}
	\displaystyle \frac{R-l^*-1}{R}&i=0,\vspace{8pt}\\
	\displaystyle \frac{R-l^*-1}{(R-i)(R-i-1)}&1\leq i\leq l^*,\\
	0&l^*<i\leq R;
\end{cases}\text{~~~~~and~~~~~}
d^*_j=\begin{cases}
	0 & j=0,\\
	\displaystyle \frac{1}{R-j}& j<l^*,\\
	\displaystyle 1-\sum_{i=0}^{l^*-1}\frac{1}{R-i}& j=l^*,\\
	0& l^*<j \leq R.
\end{cases}\]

The equilibrium for in-between cases, i.e. $0<t<1$, depends on the magnitude of $t$. For $\displaystyle t<\frac{1}{R-l^*}$, we get the same equilibrium mix as in Theorem \ref{theorem:equilibrium}. For higher values of $t$, the equilibrium mixes switch to those presented above. We note that for $0<\displaystyle t<\frac{1}{R-l^*}$, the expected payoff for Attacker actually increases by a small amount to $(1+t)R$. Asymptotically, this $t-$value range vanishes along with these gains. In sum, the precise form of the tie-breaking rule has no substantial impact on the essential features of the model.

\subsection{Risk}

\begin{figure}[ht!]
	\center
	\includegraphics{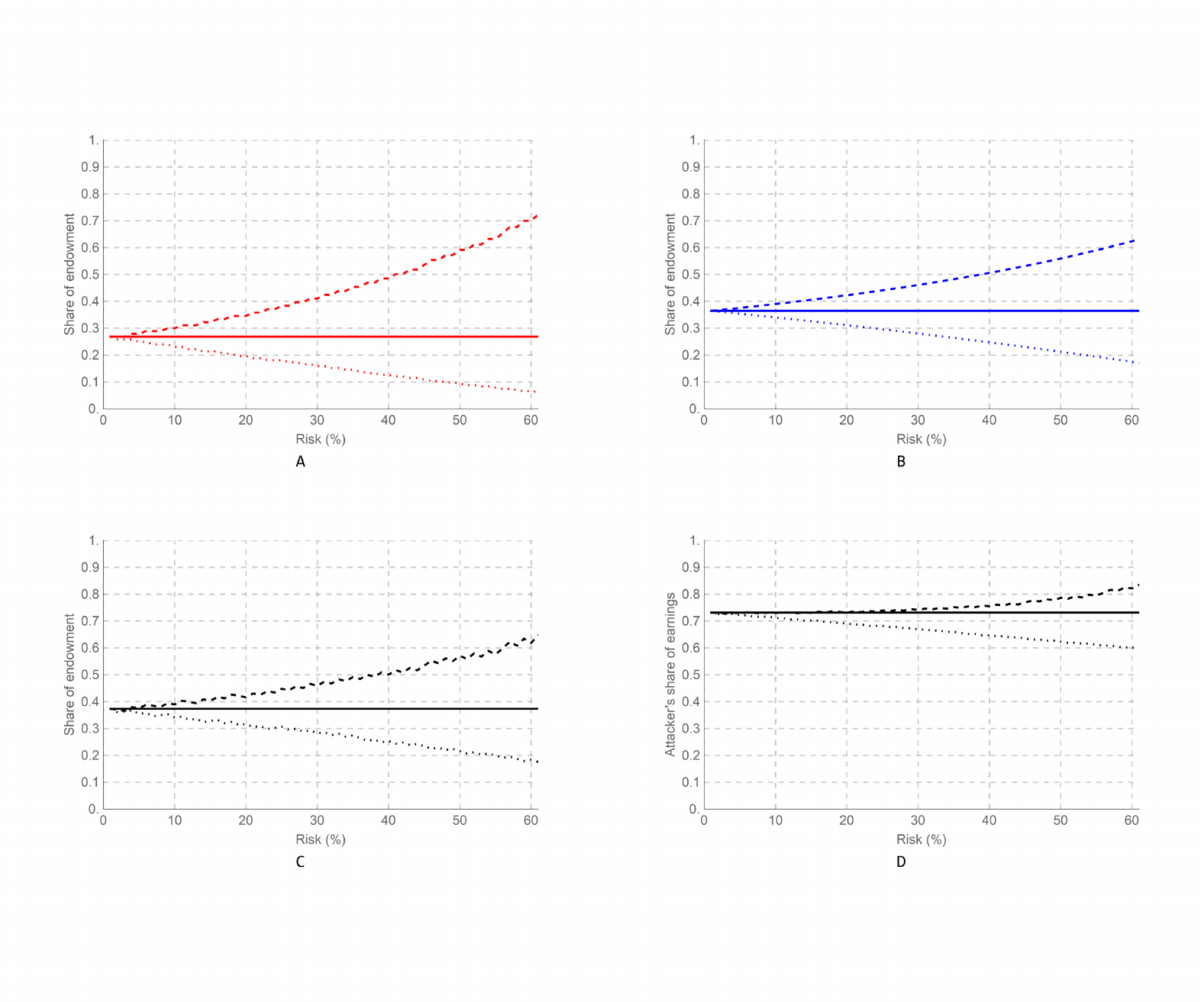}
	\caption{The effects of risk on the equilibrium, depending on what it impacts: only Attacker's uninvested endowment (dashed line), only the value appropriated by Attacker when winning (dotted line), or neither/both at the same time (solid line). Displayed equilibria were calculated for equal endowments of $100$. The following aspects of equilibrium are depicted: average Attacker investment ({\bf A}), average Defender investment ({\bf B}), probability for Attacker to win ({\bf C}), and Attacker's expected share of total earnings ({\bf D}).}\label{fig:risk}
\end{figure}

The model presented in the opening of Section \ref{section:equal} was predicated on the assumption that payoffs occur with certainty. As in symmetric contests, payoffs can be probabilistic, in the sense that a random event may ex post wipe out some of the players' payoffs. Examples of such probabilistic events include climate shocks, economic developments and political instabilities \citep{coronese19, mach19, duncan72, baker16}. In our model with players' roles being fixed ex ante, such events that occur with a fixed probability $p$ can apply to any/all of the following:

\begin{enumerate}
	\item[a] Attacker's uninvested endowment;
	\item[b] the value appropriated by Attacker;
	\item[c] Defender's endowment.
\end{enumerate} 
Option {\bf c}~amounts to a mere rescaling of Defender's payoffs, and thus has no effect on the equilibrium. The case when all of Attacker's potential earnings are under risk, i.e., both {\bf a} and {\bf b} are simultaneously affected by the event is analogous. We may thus compare three scenarios: no risk (i.e., the setting discussed thus far), risky remaining endowments and risky appropriation. The expected proportional conflict investments, win rates, and relative earnings are shown in Fig. \ref{fig:risk}. The overall result is unsurprising: risky remaining endowments lead to enhanced, while risky appropriation leads to dampened conflict.

\section{Unequal endowments}\label{section:unequal}

Although asymmetric conflict can involve players with equal power, conflict often emerges between players differing ex ante in power. Existing literatures on contests, rent-seeking and (all-pay) auctions consider symmetric games of strategy. Accordingly, the manner in which power differences shape attack and defense remains an open question \citep{dreu19}. On the one hand, Attackers with a lower initial endowment may be less inclined to invest in conflict, effectively reinforcing the ex ante power differences. Vice versa, those with a higher endowment than their Defender would be more inclined to aggress and thereby amplify wealth and power differences between Attacker and Defender \citep[viz. Conflict Spiral Theory; ][]{dreu95, deutsch73, kydd00, abbink21, bacharach81}. On the other hand, Attackers with a lower endowment have relatively much to gain from winning the contest. Hence, having less power than Defender may lead to a more rather than a less forceful attack (\citealt{hirshleifer91, skaperdas92}; also see \citealt{dreu21}).

Allowing for possibly asymmetric starting positions for the two sides leads to a natural generalization of the model presented in the previous section. Specifically, when Attacker (Defender) begins with an endowment of $R_A\geq 2$ (respectively, $R_D\geq 2$), their respective strategy sets change correspondingly, and the payoff functions become:
\[\pi_A(i,j)=
\begin{cases}
	R_A-i+R_D-j & i>j,\\
	R_A-i& i\leq j;
\end{cases}\text{~~~~~and~~~~~}
\pi_D(i,j)=
\begin{cases}
	0 & i>j,\\
	R_D-j& i\leq j.
\end{cases}\]

Take $\bar{R}$ to be the larger of $R_A$ and $R_D$, and note that in equilibrium, the `richer' side will never contribute more than $\bar{R}$ to conflict: If Attacker is richer, an attack stronger than $\bar{R}$ guarantees a win, yet turns a loss into a victory only when there is nothing left to appropriate from Defender. In case Defender is richer, defending with $\bar{R}$ already guarantees successful defense, and thus higher contributions on their part would be wasteful.

\subsection{Equilibrium}\label{section:uneq:equilibrium}

\begin{theorem}\label{uneq:equilibrium}
	The A-D game with endowments of $R_A, R_D$ has a single mixed-strategy Nash equilibrium $(a^*,d^*)$, with:
	\[a^*_i=\begin{cases}
		\displaystyle \frac{R_D-l^*}{R_D}& i=0,\\
		\displaystyle \frac{R_D-l^*}{(R_D-i+1)(R_D-i)}& 1\leq i\leq l^*,\\
		0& l^*<i\leq R_A;
	\end{cases}\text{~~~~~and~~~~~}
	d^*_j=\begin{cases}
		\displaystyle \frac{1}{R_D-j}& j<l^*,\\
		\displaystyle 1-\sum_{j=0}^{l^*-1}\frac{1}{R_D-j}& j=l^*,\\
		0& l^*<j\leq R_D.
	\end{cases}\]
	The value $l^*$ is given by $l^*=\min(R_A,k^*)$, with $k^*$ identified by: \[H_{R_D}-H_{R_D-k^*} < 1 < H_{R_D}-H_{R_D-k^*-1}.\]
\end{theorem}

\begin{corollary}\label{uneq:total-earnings}
	The sum of expected payoffs is equal to total endowments less the upper bound
	\[\Pi_A+\Pi_D=R_A-E(a^*)+R_D-E(d^*)=R_A+R_D - l^*.\]	
\end{corollary}

\begin{corollary}\label{uneq:conflict-investments}
	The sum of expected conflict investments is equal to the upper bound \[E(a^*)+E(d^*)=l^*.\]
\end{corollary}

Looking at the asymptotic properties, we can ask what happens to the equilibrium when either endowment grows infinite while the other one stays fixed. It follows from Theorem \ref{uneq:equilibrium} that as $R_A$ grows, its magnitude becomes irrelevant to the equilibrium structure, since whenever $R_A > R_D$, all the variables of importance depend only on $R_D$. In the other case with $R_D > R_A$, we get $l^*=R_A$. Here, the asymptotic properties are more interesting:

\begin{theorem}\label{uneq:upper-bound-limit-attacker} 
As Defender's endowment goes to infinity while Attacker's endowment is fixed, the following hold:
\begin{itemize}
	\item $\lim_{R_D\to \infty} d^*_{R_A}=1.$
	\item $\lim_{R_D\to \infty} a^*_{0}=1.$
	\item $\lim_{R_D\to \infty} P(W)=0.$	
\end{itemize}
\end{theorem}

\subsection{Interpretation}

A number of similarities between the cases of equal and unequal endowments are manifest. The circular best-response logic described in Section \ref{eq:interpretation} applies in the unequal scenario, too. Crucially, we get a unique equilibrium in mixed strategies in this setting, too. 

Concerning the specific structure of the equilibrium, the upper limit $l^*$ again plays an important role. Recall that this upper limit is primarily determined by Defender's endowment. Theorem \ref{uneq:equilibrium} indicates that whenever $R_A >R_D$, the equilibrium mixes are precisely the same as when both players are equally endowed with $R_D$. Thus, compared with the equal power case, a more powerful Attacker will not be more aggressive, nor win more frequently. 

The case of a more powerful Defender appears more interesting. We have shown that the upper limit here is determined by $l^*=\min(R_A,k^*)$, and $k^*$ is the value at which the `inverted harmonic series' starting at $\frac{1}{R_D}$ crosses the threshold of $1$. Defender's equilibrium strategy thus changes in shape only whenever $R_A<k^*$, i.e., when Attacker is at least $\approx 37\%$ poorer. As Attacker's endowment drops below this value, the probability weight assigned to defending at the upper limit acquires an ever larger weight, converging to a value of $1$ as the share of Defender's initial endowment approaches $100\%$, see Theorem \ref{uneq:upper-bound-limit-attacker}. 

\begin{figure}[ht!]
	\center
	\includegraphics{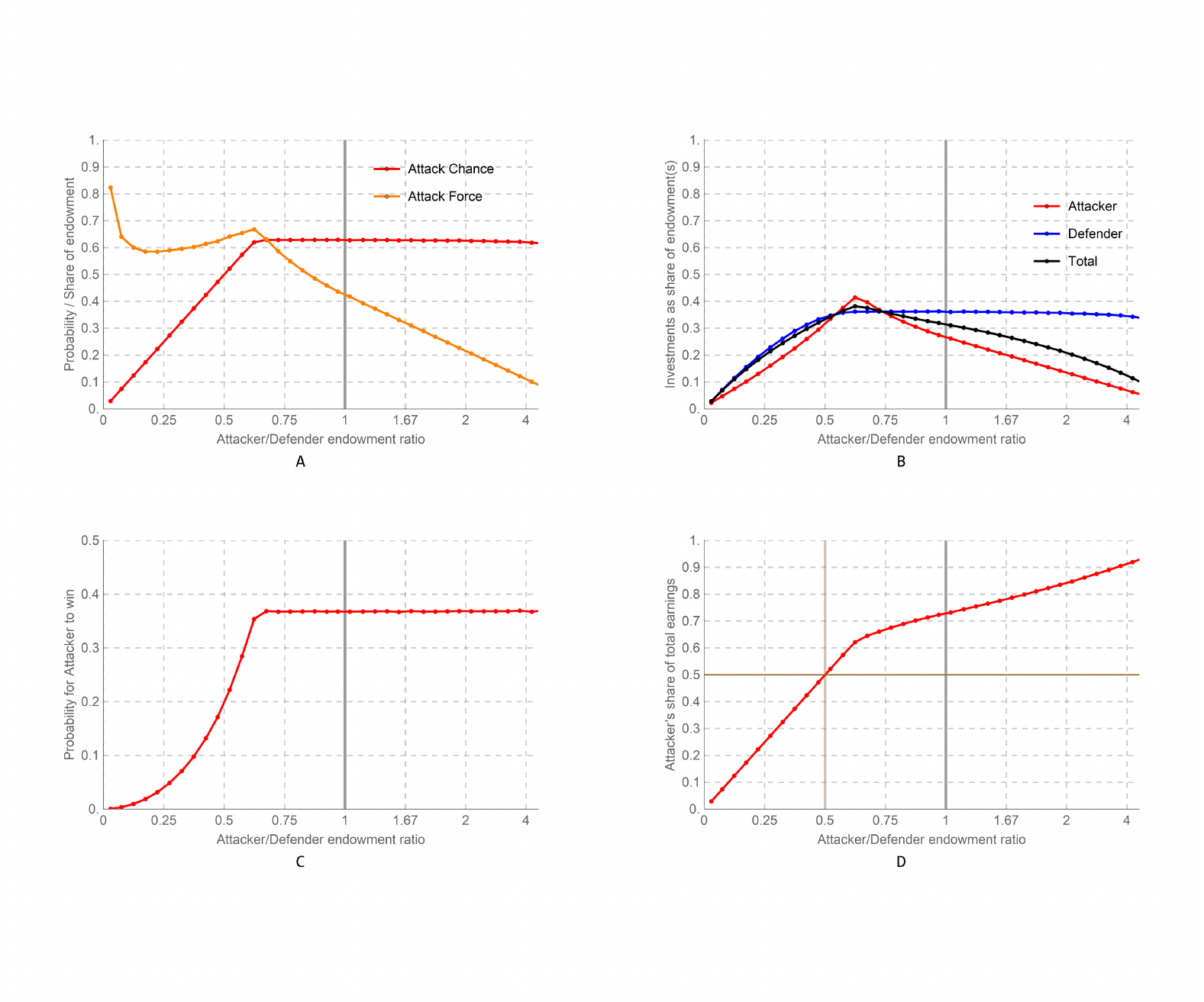}
	\caption{Properties of the equilibrium under various initial endowment ratios. {\bf A}: Probability for Attacker to choose a  positive conflict investment (red), and attack force (orange). (Attack force is the expected conflict investment for Attacker conditional on choosing $a>0$.) {\bf B}: Expected equilibrium investments as share of initial endowment for Attacker (red), Defender (blue), expected total investments as share of the sum of endowments (black). {\bf C}: Probablity for Attacker to win. {\bf D}: Average Attacker earnings as share of total earnings. The intersection of the brown lines exemplifies the `paradox of power': Attacker starting with half as many resources as their counterpart nevertheless has the same expected earnings. In all panels, vertical black lines indicate equal endowments.}\label{fig:uneq_combined}
\end{figure}

Intuitively, Attacker facing a much richer Defender ($R_A<k^*$) would be inclined to attack more frequently, and devote more to attack (i.e., along the `paradox of power'). This intuition, however, turns out to be incorrect, since $a^*_0=1-\frac{R_A}{R_D}$ is increasing in $R_A$, see also Fig.~\ref{fig:uneq_combined}A. We saw in Section \ref{section:equal} that Defender invests, on average, more in conflict. In settings with unequal power, this is still mostly true. However, a range in which Attacker invests a larger share of their endowment now emerges, specifically when Attacker's endowment is between $50\%$ and $25\%$ less than Defender's, see Fig. \ref{fig:uneq_combined}B.

In terms of win rates for Attacker, the condition of $l^*=\min(R_A,k^*)$ splits the endowment domain in two. Whenever Attacker is rich enough such that $l^*=k^*$, it wins $\approx 37\%$ of interactions, just like under equality. On the other hand, if $R_A$ drops below $k^*$, we saw that Defender will be ever more likely to choose the maximal reasonable defense of $R_A$, and this translates to decreasing Attacker's chances to win, see Fig.~\ref{fig:uneq_combined}C. At the limit of inequality favoring Defender -- consistent with our intuition -- Attacker will almost never win.

In terms of the relationship between ex ante and ex post inequality, Fig.~\ref{fig:uneq_combined}D indicates a segmental linear relationship. In the first segment, characterized by $R_A<k^*$, Attackers catch up to Defenders rather quickly. For instance, with a $50\%$ lower endowment, Attackers manage to equalize the ex ante wealth differences, exemplifying the `paradox of power' \citep{hirshleifer91, durham98}. This equalization can be again attributed to the high probabilities with which Defenders choose their maximum reasonable defensive action. Along the second segment, in which outcomes favor Attackers, the growth in inequality can be more directly ascribed to the growth in Attacker's initial endowment, rather than to the investment needed by Defender to play their equilibrium strategy.

Because conflict induces efficiency loss on the social level, policy makers may aim to reduce conflict magnitude. Our results show that conflict leads to the largest proportional waste at $\approx 63\%$ relative endowment for Attacker. This means that social inefficiency can be reduced by making either side poorer or richer. Examples abound for each of these options, in the forms of international embargoes and sanctions that impoverish Attackers, development aid that enrich Attackers, and defense assistance that enrich Defenders. Crucially, therefore, conflict can be reduced both by making attackers stronger and/or making defenders weaker. This resonates with Power Balance and Deterrence Theories \citep{bacharach81, darcy11, gartzke09} in that (symmetric) conflict is less likely in the presence of power balance rather than imbalance. The argument is that with power imbalance, the stronger player becomes increasingly less deterred from aggressing its weaker antagonist. Our model suggests a similar pattern, while ascribing it to a distinctly different mechanism. Rather than the stronger player becoming less deterred, in our model it is the weaker Attacker who becomes, up to a certain point, more interested in acquiring Defenders' endowment, and hence more aggressive.

\subsection{Warm-glow and harm aversion}

\begin{figure}[ht!]
	\center
	\includegraphics{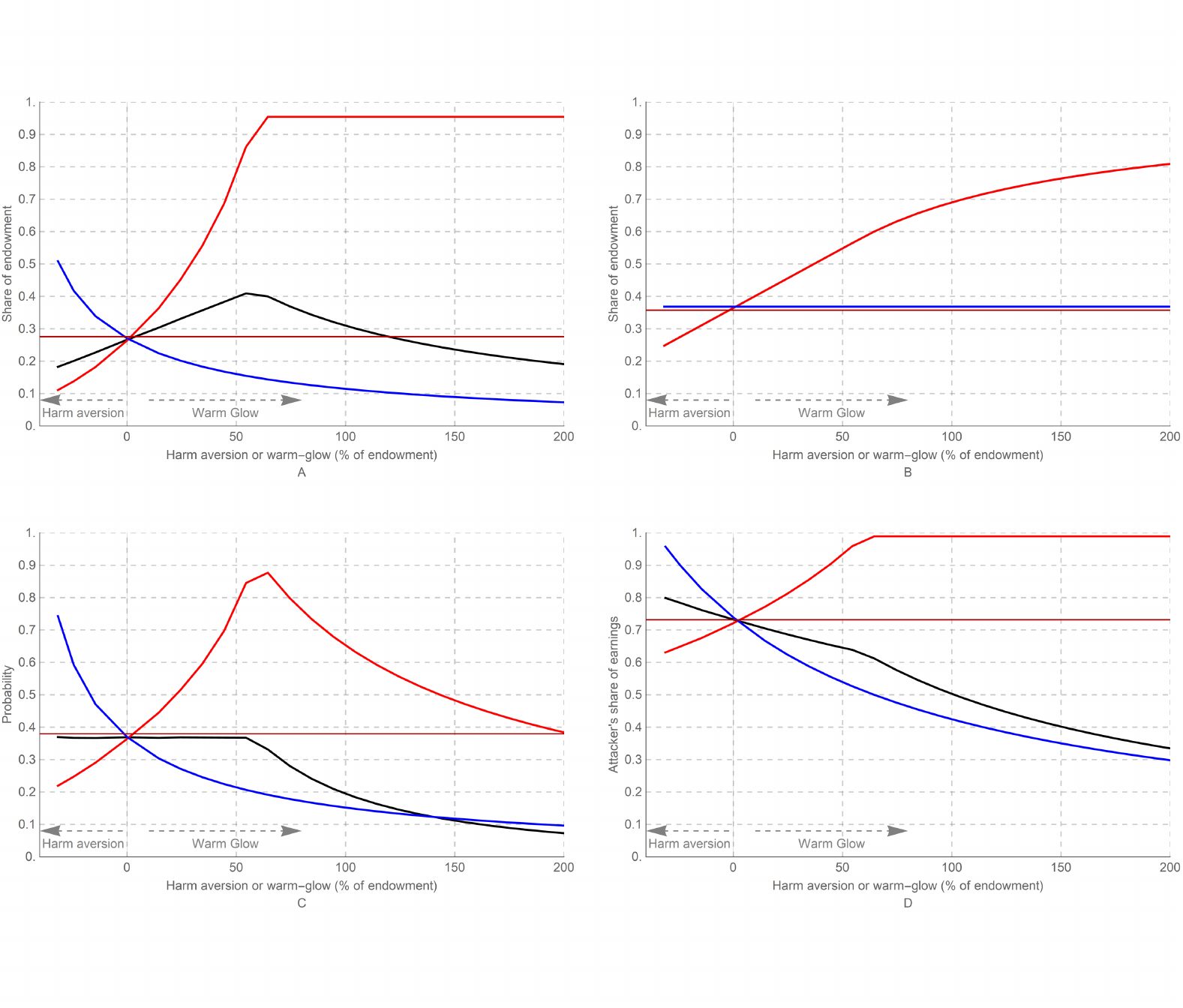}
	\caption{The effects of warm-glow of winning / harm aversion  at its various levels depending on which players it affects: only Attacker (red), only Defender (blue), both players (black), or neither player (brown). Displayed equilibria were calculated for equal endowments of $100$. The following aspects of equilibrium are depicted: average Attacker investment ({\bf A}), average Defender investment ({\bf B}), probability for Attacker to win ({\bf C}), and Attacker's expected share of total earnings ({\bf D}). On panel {\bf B}, the  red and black curves coincide, as do the blue and brown ones. Note: In order to get a unique equilibrium, a minimum warm-glow value of $1$ was set.}.\label{fig:warmglow_both}
\end{figure}

A common observation in the empirical conflict literature is that players deviate from the Nash equilibrium. One the one hand, players should underinvest in conflict when they hold social preferences and care about each other's outcomes and want to avoid causing harm \citep[e.g., ][]{fehr99, bolton00, rabin93, charness02}. Such social preferences can be theoretically captured by assuming that actors gain extra utility by not harming others \citep{dreu19b}. This should pertain especially if not exclusively to Attackers, with the straightforward result that the presence of harm aversion should reduce conflict effort among Attackers more than among Defenders, see Fig. \ref{fig:warmglow_both}.

On the other hand, players frequently overinvest in contests \citep{chowdhury14, dechenaux15}. Such over-investment suggests that human actors gain utility from winning a conflict \citep{rojek20, ku05}, i.e., a `warm-glow' of victory. If we assume a warm-glow for both actors, a layer of symmetric conflict (i.e., competition for the warm-glow prize of winning), similar to a standard Tullock contest, is added to the asymmetric underlying Attacker-Defender conflict.

When players start with equal endowments, the consequences of warm-glow depend on their magnitude (c.f., Fig. \ref{fig:warmglow_both}, black lines). While Defender's investment increases monotonically with the magnitude of warm-glow (Fig. \ref{fig:warmglow_both}B), Attacker's investment increases initially, until the magnitude of warm-glow reaches $~50\%$ of the initial endowment; and decreases thereafter, falling even below the level of no warm-glow(\ref{fig:warmglow_both}A). Overall, despite the larger investments, Attacker is unable to increase their chances to win, even in the initial range where their conflict investment relates positively to the magnitude of warm-glow. Defender's high investments succeed in demotivating Attacker, who will win the conflict ever more rarely at higher levels of warm-glow (Fig. \ref{fig:warmglow_both}C).

Sometimes, warm-glow is experienced by one contestant, but not (or less so) by the other. The equilibria of such games are combinations of the distributions described in Sections \ref{subsection:equilibrium} and \ref{section:uneq:equilibrium}, depending on which side receives utility from winning. For example, if only Defender experiences warm-glow, Attacker's probability distribution follows the formula in Section \ref{subsection:equilibrium}. In this case, Defender's equilibrium distribution follows the formula in Section \ref{subsection:equilibrium}, assuming $RD=R+w$, where $R$ is the initial endowment, and $w$ is the value of warm-glow of winning.

In Fig. \ref{fig:warmglow_both}, we compare all four possible scenarios with warm-glow affecting none, only one, or both players. As expected, we observe the highest (lowest) levels of conflict and ex post inequality when only Attacker(Defender) is affected by warm-glow. Fig. \ref{fig:warmglow_both}B shows that Defender's equilibrium behavior depends only on whether Attacker experiences warm-glow. Interestingly, we see that beyond about two-thirds of the players' endowment, further increase in the magnitude of warm-glow has little effect on the equilibrium behavior, and typically diminishes Attacker's chances to win.

\section{Outside options}\label{section:outside}

Thus far we considered Attack-Defense conflicts in which Attackers' rent-seeking attempts are their only possibility to increase their wealth. Extant work on the `guns versus butter' problem highlights, however, that actors oftentimes have multiple options to accumulate wealth, with only one being predatory capture \citep{skaperdas01, carter21, powell93, russett70}. Along similar lines, Bargaining theory has shown how `outside options' can alter players' competitiveness \citep{compte02, kahn93, hennigschmidt18}. This raises the question how having `non-conflict' alternatives for wealth accumulation (henceforth, `production') alters the dynamics and outcomes of Attack-Defense games. 

\subsection{Equilibrium}

The cost of production is set at $2\leq C \leq \min(R_A,R_D)$, and the benefit is set at some value $B>C$. The net surplus from production will be denoted by $Q=B-C$. We assume production to be risk-free and Defender's benefits of production to be under risk of appropriation in case of a successful attack. Endowments are equal to $R_A$ and $R_D$, respectively. The strategy sets can be taken to be unchanged. The payoffs are given by:

\[\pi_A(i,j)=
\begin{cases}
	R_A-i+R_D-j+2Q & i>j\text{~and~}R_A-i\geq C\text{~and~} R_D-j\geq C,\\
	R_A-i+R_D-j+Q & i>j\text{~and~} R_A-i<C\text{~and~} R_D-j\geq C,\\
	R_A-i+R_D-j & i>j\text{~and~}R-i<C\text{~and~}R-j<C,\\	
	R_A-i+Q & i\leq j\text{~and~}R_A-i\geq C,\\
	R_A-i& i\leq j\text{~and~}R_A-i<C;
\end{cases}
\]
and
\[
\pi_D(i,j)=
\begin{cases}
	R_D-j+Q& i\leq j\text{~and~}R_D-j\geq C,\\
	R_D-j& i\leq j\text{~and~}R_D-j<C,\\
	0 & i>j.
\end{cases}\]	

\begin{theorem}
With cost $C\geq 2$ and benefit $B>C$ (and corresponding profit of $Q=B-C$), the A-D game with endowments of $R_A$ and $R_D$ has a single, mixed-strategy Nash equilibrium $(a^*,d^*)$.
\end{theorem}

The formulas describing the equilibrium strategies are specified in the Appendix.

\subsection{Interpretation}\label{subsec:outside:interpretation}

\begin{figure}[ht!]
	\center
	\includegraphics{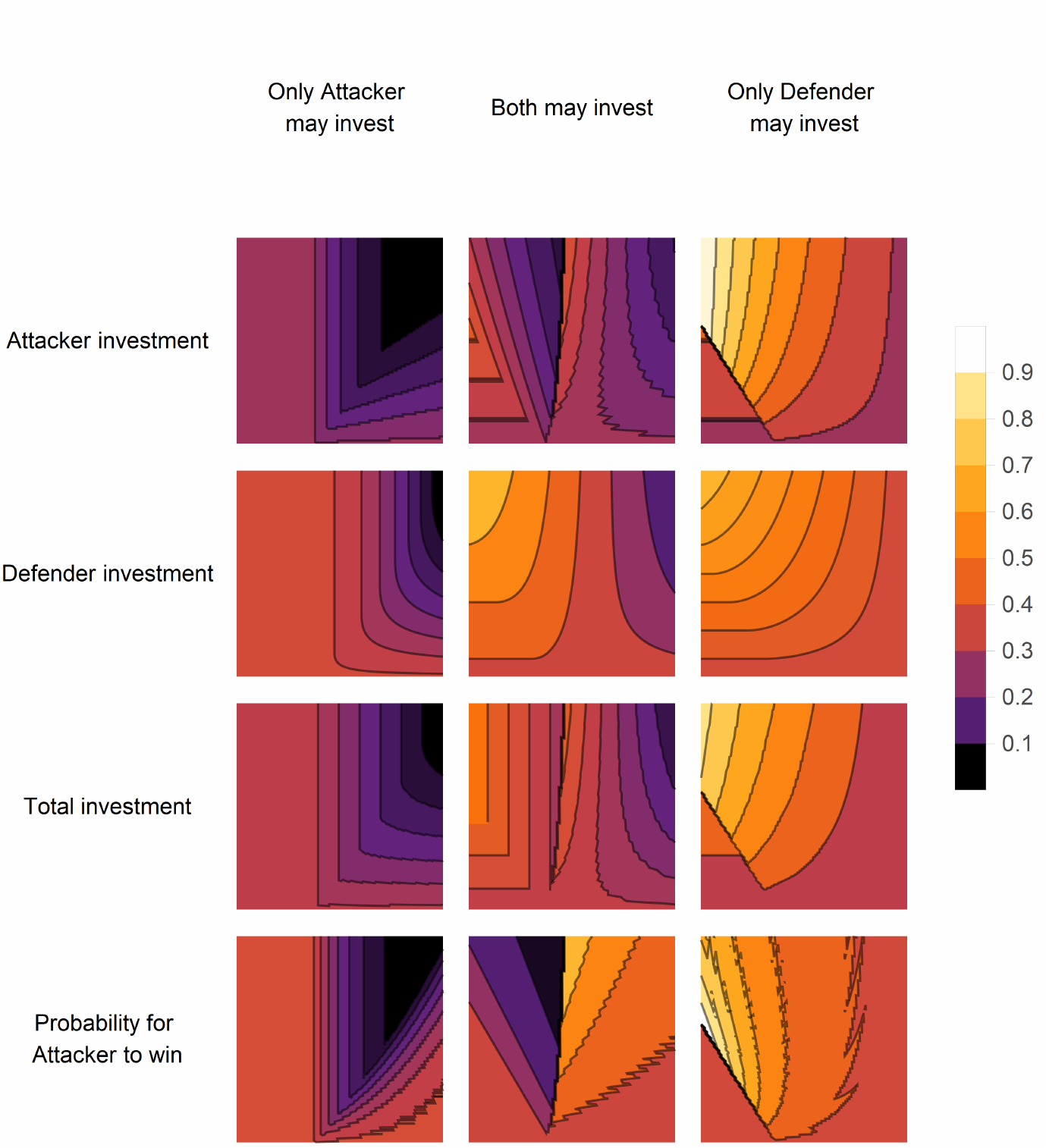}
	\caption{Equilibrium values of various parameters at diverse values of cost ($C$, $x$-axes) and profit ($\Pi$, $y$-axes). Various scenarios are depicted in the columns, as follows. Left column: only Attacker has a production option. Center column: both players have a production option (with identical parameters). Right column: only Defender has a production option. Variables of interest are organized in rows. Row $1$: Attacker's investment as share of the endowment. Row $2$: Defender's investment as share of the endowment. Row $3$: Total investment as share of the endowments, i.e., social waste. Row $4$: Probability for Attacker to win. Displayed equilibria were calculated for equal endowments of $100$.}\label{fig:production}
\end{figure}

Adding production options significantly complicates the A-D game. Even treating `rescaled' versions of the game as equivalent, we face a four-dimensional parameter space. To keep the discussion tractable and visualization feasible, we limit ourselves here to the case of equal endowments. We compare settings in which only Attacker, both players, or only Defender has a production option.

When only Attacker may invest productively (see Fig.~\ref{fig:production}, panels A1-A4), the structure of conflict is unchanged when production costs are low. At higher costs, conflict tends to be reduced, especially when the net benefits of production are high. Thus, providing Attackers with high-cost production options with appealing benefits is an effective way to mitigate conflict.

In contrast, if only Defender has a production option (Fig.~\ref{fig:production}, panels C1-C4), we see an {\it increase} in hostilities, especially when the costs of production are low, and the benefits high. This resonates with the literature on the `natural resource curse': countries who have cheap access to large economic benefits become particularly attractive targets for hostile takeovers \citep{ploeg11, brunnschweiler09}, and much of these benefits will need to be spent on defensive efforts. Our analysis shows that even spending a large share of the endowment on defense turns out to be ineffective. At low costs, the crossing of a definite threshold characterizes the transition from `normal' to extremely high levels of conflict (panels C1, C3, and C4).

When both players have identical production options, the parameter spaces exhibits surprising non-monotonicities in both the cost and the benefit parameter, and in the threshold transitions between low and high conflict (see especially panels B1 and B4). In such situations, the effectiveness of interventions is difficult to anticipate, especially at intermediate levels of production cost. The natural resource curse reappears (panels B1-B2), although Defenders are now better able to counteract Attacker hostility.

Overall, providing Attacker with a costly, high-yield production option emerges as an effective way to reduce conflict. Subsidizing Defender appears, however, highly counter-effective. Richer Defenders, whether because of ex ante windfalls or due to having outside production facilities, are likely targets of predator attack and aggression.

\section{Conclusion and implications}\label{section:conclusion}

Asymmetric Attacker-Defender conflicts, wherein one party attempts to gain what the other party has, have an internal logic that sets them apart from symmetrical role conflicts. By investing energy in defense, predation can be averted but overall welfare is reduced. Examples like scorched-earth tactics illustrate how wasteful such conflicts can be, and overinvestment on both sides can create the proverbial `Pyrrhic victory' in which both parties are worse off and nothing is gained, since the prize one party tries to defend, and the other party tries to obtain is destroyed in the process. Indeed, when the predatory and defending parties are approximately equally endowed, and outside options are irrelevant, our `$1/E$-rules' provide a useful rule of thumb on what to expect. Equilibrium Attackers will withhold from predatory attempts $1/E$ of the time; and will also succeed in $1/E$ of cases. Collectively, about one-third of all resources will be wasted on conflict.

Unequal endowments change both the space of possibilities and the incentives for attack and defense. Richer Attackers will merely regard their power advantage as a `pure surplus', and the equilibrium strategies essentially remain unchanged. Poor Attackers, however, may increase their hostility, especially when potential predatory gains are combined with realistic chances for victory. We showed this to be most likely when Attackers start out with around one-third less resources than Defenders.

We pointed out several factors that can reduce conflict. First, various sources of risk impact the tradeoff between conflictual and non-conflictual earnings. For example, Attacker invest less in conflict the more risk is placed on predatory gains. Second, increasing harm aversion, and lowering the magnitude of `warm-glow' will also reduce the intensity of conflict, suggesting that social preferences resulting from social norms on not harming others can reduce conflict and conflict expenditures. Third, changing the relative starting endowments of the conflicting parties through, e.g., subsidies or sanctions, may be an efficient tool to mitigate asymmetric conflicts. However, aiding defending parties can create a `natural resource curse', where value provided at no cost to a Defender merely provides additional incentives for Attackers to invest energy in predation. In general, the interplay between outside options for economic growth, their costs and benefits, and conflict intensity are marked by non-monotonic relationships and critical transitions. These complexities notwithstanding, profitable alternative opportunities for predators create a substitute for conflict that, if costly enough, can move investments from rent-seeking to non-conflictual means of wealth accumulation.

Our models allow for several directions of generalization concerning the number of players, and the dynamics of their interaction. Of particular interest is that our model assumed unitary players, akin to individual decision-makers or dictators in group settings. Yet asymmetric conflicts of attack and defense often emerge between heterogeneous groups with individual members having some discretion over whether and how much to contribute to the group-level attack and defense \citep{dreu19}. The logic and equilibrium properties of our two-player model provide a framework for developing theory on such two-level strategic situations where individuals face a first-level public goods dilemma within their group, and a second-level asymmetric Attack-Defense conflict between groups. Such extension can further help to understand how political leaders and institutions can intervene to reduce effort spend on predation and protection, and reduce the waste of conflict.

\newpage

\section*{Appendix: Proofs}\label{appendix}
\addcontentsline{toc}{section}{Appendix: Proofs}

\setcounter{theorem}{0}

In the paper, we consider several versions of the Attacker-Defender Game: equal endowments (Section \ref{section:equal}), unequal endowments (Section \ref{section:unequal}), and outside options (Section \ref{section:outside}). The following lemmas hold true for each of these versions of the game, to which we refer collectively as the `A-D game'. 

Note that for the equal-endowment version, we set $R_A=R_D=R$. The harmonic series up to $i$ will be denoted by $H_i$: \[H_i=1+\frac{1}{2}+\frac{1}{3}+\dots+\frac{1}{i}.\]

\begin{lemma}
	\label{lemma:only-mixed}
	The A-D game allows for no pure-strategy equilibria.
\end{lemma}

\begin{proof}
	Suppose there is a pure-strategy equilibrium $(a^*,d^*)$, with $a^*\in S_A, d^*\in S_D$. 
	\begin{description}
		\item[Case $1$]: $a^*>d^*$.\\
		Attacker wins; thus, Defender's earnings are $0$. Consider Defender choosing the pure strategy $d'=R_D-1$. If $a^*<d'$, then Defender is strictly better off by choosing $d'$. If $a^*>d'$, then Attacker will capture just $1$ unit of payoff from Defender, at an expense of at least $R_D=2$, which is clearly dominated by $a'=0$.
		\item[Case $2$]: $0 < a^*\leq d^*$.\\  
		Attacker loses in this situation, and so would be strictly better off by choosing $a'=0$ rather than $a^*$.
		\item [Case $3$]: $a^* = 0$.\\
		Defender's best-response is $d^*=0$. But then Attacker is strictly better off by choosing $a'=1$, and, thereby, capturing $\leq 2$ units of payoff.
		
	\end{description}
\end{proof}

\begin{lemma}\label{lemma:suppa=suppd1} In any mixed-strategy equilibrium of the A-D game, the supports of the equilibrium strategies coincide, or differ only by one element. Specifically: $supp(a^*)=supp(d^*)$ or $supp(a^*)=supp(d^*)\cup\{l^*+1\}$, with $l^*=\max supp(d^*)$. 
\end{lemma}

\begin{proof}
	First, we show that $supp(d^*)\subseteq supp(a^*)$. Suppose the contrary, and take the largest $i$ in $supp(d^*)$ with $i\not \in supp(a^*)$. If $i=0$, we get $\pi_D(a^*,d^*)=\pi_D(a^*,0)=0$, so $d^*$ is clearly not a best-response strategy for Defender. If instead $i>0$, then $i-1$ successfully defends against the same strategies as $i$, and thus $\pi_D(a^*,i-1)>\pi_D(a^*,i)$. This means $i$ cannot be in $supp(d^*)$ after all.
	
	Next, we show that either $supp(a^*)\subseteq supp(d^*)$, or $supp(a^*)\subseteq supp(d^*)\cup\{l^*+1\}$. Suppose the contrary, and take any index $i\in supp(a^*)$ with $i\not \in supp(d^*)$. There are two possibilities: $i<l^*$, or $i>l^*$, with $l^*=\max supp(d^*)$. 
	
	In the first case, take the smallest $j\in supp(d^*)$ such that $j>i$. By the first part of the proof, $j\in supp(a^*)$. Yet, Attacker wins by choosing $j$ in the exact same cases as when choosing $i$, so $\pi_A(j,d^*)<\pi_A(i,d^*)$, meaning that $i$ can not be in $supp(a^*)$. 
	
	The other possibility is $i>l^*$. In this case, we must have $i=l^*+1$, because choosing this strategy already guarantees Attacker victory, and higher attack contributions would be wasteful.
\end{proof}

We call $l^*$ the upper bound (of equilibrium conflict investments).

\begin{lemma} In any mixed-strategy equilibrium of the A-D game, Attacker refrains from conflict with positive probability: $0\in supp(a^*)$.
\end{lemma}
\begin{proof}
	Suppose the contrary, and take $i=\min supp(a^*)>0$. Then, by Lemma \ref{lemma:suppa=suppd1}, Defender has also $\min supp(d^*)=0$. This means that Attacker never wins when choosing $i$; yet, Attacker wastes $i$ resources on attacking, implying that $\pi_A(i,d^*)<\pi_A(0,d^*)$. This shows that $a^*$ cannot be an equilibrium strategy.
\end{proof}

\begin{lemma}
	In any mixed-strategy equilibrium of the A-D game, Defender's support contains all indices up to the upper bound $l^*$: $supp(d^*)=\{0,1,\dots,l^*\}$.
\end{lemma}
\begin{proof}
	Suppose the contrary. Then, there must exist an index $i\in supp(d^*)$ with $i-1 \not \in supp(d^*)$. By Lemma \ref{lemma:suppa=suppd1}, we also have $i\in supp(a^*)$, and $i-1\not \in supp(a^*)$. But $i-1$ wins exactly in those cases in which $i$, thus $\pi_A(i-1,d^*)>\pi_A(i,d^*)$, so $a^*$ cannot be a best response to $d^*$.
\end{proof}

\begin{theorem}
	The A-D game with equal endowments of $R$ has a single, mixed-strategy Nash equilibrium $(a^*,d^*)$, with:
	\[a^*_i=\begin{cases}
		\displaystyle \frac{R-l^*}{R}&i=0,\\
		\displaystyle \frac{R-l^*}{(R-i+1)(R-i)}&1\leq i\leq l^*,\\
		0&l^*<i\leq R;
	\end{cases}\text{~~~~~and~~~~~}
	d^*_j=\begin{cases}
		\displaystyle \frac{1}{R-j}& j<l^*,\\
		\displaystyle 1-\sum_{i=0}^{l^*-1}\frac{1}{R-i}& j=l^*,\\
		0& l^*<j \leq R.
	\end{cases}\]
	
	The value $l^*$ is identified by: \[H_{R}-H_{R-l^*} < 1 < H_{R}-H_{R-l^*-1}\] that is, $l^*$ is determined by the boundary where the series $\displaystyle \frac{1}{R}+\frac{1}{R-1}+\dots$ exceeds $1$. 
\end{theorem}

\begin{proof}
	We consider Defender's equilibrium mix first.	
	
	By Lemma \ref{lemma:suppa=suppd1}. $supp(d^*)=\{0,\dots,l^*\}\subseteq supp(a^*)$. Attacker is thus indifferent between choosing $0, 1, \dots, l^*$, so $\pi_A(i+1,d^*)=\pi_A(i,d^*)$ for $i=0,\dots,l^*-1$. Using the fact that  $\pi_A(i+1,d^*)=\pi_A(i,d^*)-1+d_i(R-i)$, and solving for $d_i$ we get the indicated distribution. 
	
	Based on this result, we next prove that $supp(a^*)=supp(d^*)$. Suppose the contrary; then $supp(a^*)=supp(d^*)\cup\{l^*+1\}$. This means that the previous lemma can be extended for $i=l^*$, yielding $d_{l^*}=\frac{1}{R-l^*}$. But $d^*$ is a mixed strategy, and thus we get: \[\displaystyle \sum_{i=0}^{l^*} \frac{1}{R-i}=1 \text{ or, equivalently, } \sum_{i=R-l^*}^{R} \frac{1}{i}=1\] This is impossible, because the left-hand side is a partial sum of the harmonic series, and thus the right-hand side cannot be an integer \citep{osler12}.
	
	We can now consider Attacker's equilibrium mix. As Defender is indifferent between the pure strategies in the support of $d^*$, we have $\pi_D(a^*,i)=\pi_D(a^*,i-1)$ for $i=1,\dots,l^*$. Note that $\pi_D(a^*,0)=a_0R$. At the same time:
	\begin{align*}
		\pi_D(a^*,i)=&(1-P_W(a^*,i))(R-i)=\\
		=&(1-(P_W(a^*,i-1)-a_{i}))(R-i)=\\
		=&(1-P_W(a^*,i-1)(R-i)+a_i(R-i)=\\
		=&\pi_D(a^*,i-1)\frac{R-i}{R-(i-1)}+a_{i}(R-i).
	\end{align*}
	Using the relation $\pi_D(a^*,i)=\pi_D(a^*,i-1)$, we get:
	\[a_{i}=a_0\frac{R}{(R-i+1)(R-i)}\]
	
	Since  $supp(a^*)=supp(d^*)$, Defender always wins when defending with $l^*$, so $\pi_D(a^*,l^*)=R-l^*$. This yields $\displaystyle a_0=\frac{R-l^*}{R}$, showing that the equilibrium mix has the claimed form.
\end{proof}

\begin{corollary}
	The sum of expected payoffs is equal to the total of endowments less the upper bound:
	\[\Pi_A+\Pi_D=2R - l^*.\]	
\end{corollary}
\begin{proof}
	From Theorem \ref{theorem:equilibrium}, it is transparent that the expected earnings for Attacker is $R$, and for Defender is $R-l^*$, which directly implies the result.
\end{proof}

\begin{corollary}\label{corollary:sum-conflict-investments}
	The sum of expected conflict investments is equal to the upper bound: \[E(a^*)+E(d^*)=l^*.\]
\end{corollary}
\begin{proof}
	The initial endowment are diminished only through conflict investments, i.e., in expectation by $E(a)$ and $E(d)$. The sum of these endowments less investments is equal to sum of expected payoffs. Using Corollary \ref{total-earnings}, we have: \[2R-E(a^*)-E(d^*)=2R-l^*,\] which is equivalent to the claim.
\end{proof}

\begin{theorem} 
	In the A-D game with equal endowments of $R$, the ratio of the upper bound to endowment converges to  $\displaystyle 1-\frac{1}{e}$ as the amount of endowment goes to infinity:
	\[\displaystyle \lim_{R\to \infty} \frac{l^{*}}{R}=1-\frac{1}{e}.\]
\end{theorem}
\begin{proof}
	Recall that $l^*$ is identified by: \[H_{R}-H_{R-l^*} < 1 < H_{R}-H_{R-l^*-1}.\] From the properties of the harmonic series, we have:	
	
	\begin{alignat*}{2}
		\ln(R)+\gamma & < H_R & & < \ln(R+1)+\gamma \\
		\ln(R-l^*)+\gamma & < H_{R-l^*} && < \ln(R-l^*+1)+\gamma\\	
		-\ln(R-l^*+1)-\gamma & < -H_{R-l^*} && < -\ln(R-l^*)-\gamma\\
		-\ln(R-l^*-1+1)-\gamma & < -H_{R-l^*-1} && < -\ln(R-l^*-1)-\gamma\\
		\ln\left(\frac{R}{R-l^*+1}\right) & < H_R-H_{R-l^*} && < \ln\left(\frac{R+1}{R-l^*}\right)\\
		\ln\left(\frac{R}{R-l^*}\right) & < H_R-H_{R-l^*-1} && < \ln\left(\frac{R+1}{R-l^*-1}\right)\\
		\ln\left(\frac{R}{R-l^*+1}\right) & < 1 && < \ln\left(\frac{R+1}{R-l^*-1}\right)\\
		\ln\left(\frac{R-l^*-1}{R+1}\right) & < -1 && < \ln\left(\frac{R-l^*+1}{R}\right)\\
		\frac{l^*-1}{R} & < 1-\frac{1}{e} && <\frac{l^*+2}{R+1}.
	\end{alignat*}
	Taking limits, the claim follows.
\end{proof}

\begin{corollary}
	Attacker's share of earnings converges to $\displaystyle \frac{e}{e+1}$:
	\[\displaystyle \lim_{R\to \infty} \frac{\Pi_A}{\Pi_A+\Pi_D}=\frac{e}{e+1}.\]
\end{corollary}
\begin{proof} We have: \[\displaystyle \lim_{R\to \infty} \frac{\Pi_A}{\Pi_A+\Pi_D}=\displaystyle \lim_{R\to \infty} \frac{R}{R+R-l^*}=\frac{1}{2-1+\frac{1}{e}}=\frac{e}{e+1}.\]
\end{proof}

\begin{corollary} 
	Expected conflict investments proportional to endowment converge to $\displaystyle 1-\frac{2}{e}$ for Attacker, and $\displaystyle \frac{1}{e}$ for Defender as the endowment goes to infinity:
	\[\lim_{R\to \infty} \frac{E(a^*)}{R}=1-\frac{2}{e}\text{~~~~~and~~~~~} \lim_{R\to \infty} \frac{E(d^*)}{R}=\frac{1}{e}.\]
\end{corollary}
\begin{proof}
	\begin{align*}
		E(d^*) = & \displaystyle \sum_i^{l^*}d_ii\\
		=& d_{l^*}(R-l^*)+\sum_i^{l^*-1}\frac{1}{R-i} i\\
		= & d_{l^*}(R-l^*)-\sum_i^{l^*-1}\frac{-i}{R-i} \\= & d_{l^*}(R-l^*)-\sum_i^{l^*-1}\left(\frac{R-i}{R-i} -\frac{R}{R-i}\right)\\
		= & d_{l^*}(R-l^*)-l^*+R\sum_i^{l^*-1}\frac{1}{R-i}\\
		= & d_{l^*}(R-l^*)-l^*+R(1-d_{l^*})\\
		= & R-l^*-d_{l^*}l^*
	\end{align*}
	Clearly $\lim_{R\to \infty}d_{l^*}=0$, so we get: \[
	\lim_{R\to \infty} \frac{E(d^*)}{R} =  1-\left(1-\frac{1}{e}\right)= \frac{1}{e}.\]
	Corollary \ref{conflict-investments} implies $\displaystyle \frac{E(a^*)}{R}= \frac{l^*}{R}-\frac{E(d^*)}{R}$. Using Theorem \ref{upper-bound-limit}, we get \[\lim_{R\to \infty} \frac{E(a^*)}{R}=1-\frac{2}{e}.\]
\end{proof}

\begin{corollary} Defender's expected conflict investment proportional to endowment when Attacker wins (loses) converges to $3-e$ $\left(\text{respectively, } \displaystyle1-\frac{1}{e-1}\right)$:
	\[\displaystyle \lim_{R\to \infty} \frac{E(d|W)}{R}=3-e \text{~~~~~and~~~~~}
	\lim_{R\to \infty} \frac{E(d|L)}{R}=1-\frac{1}{e-1}.\]
\end{corollary}
\begin{proof}
	Attacker earns an expected payoff determined by expected conflict expenditures and gains, i.e., \[\Pi_A=R-E(a)+P(W)(R-E(d|W)).\]
	At the same time, we have $\Pi_A=R$, and thus, after reorganizing terms, we get $\displaystyle E(d|W)=R-\frac{E(a)}{P(W)}$. Using Corollary \ref{conflict-investments-limit}, we get \[\displaystyle \lim_{R\to \infty} \frac{E(d|W)}{R}=1-\frac{1-\frac{2}{e}}{\lim_{R\to \infty}P(W)}.\]
	
	Using the law of total probability, the expected expenditure of Defender is given by:
	\[E(d)=P(W)E(d|W)+(1-P(W))E(d|L)\]
	
	Consider Defender first, whose expected payoff of $R-l^*$ is also equal to $(1-P(W))(R-E(d|L))$, i.e., the probability of Attacker losing times the expected amount of remaining endowment. 
\end{proof}

\begin{theorem}
	The probability of Attacker winning converges to $\displaystyle \frac{1}{e}$ as the endowment goes to infinity:
	\[\displaystyle \lim_{R\to \infty} P(W)=\frac{1}{e}.\]
\end{theorem}
\begin{proof}
	The theorem follows directly from combining the results of the previous three corollaries.
\end{proof}

\begin{theorem}
	The A-D game with endowments of $R_A, R_D$ has a single mixed-strategy Nash equilibrium $(a^*,d^*)$, with:
	\[a^*_i=\begin{cases}
		\displaystyle \frac{R_D-l^*}{R_D}& i=0,\\
		\displaystyle \frac{R_D-l^*}{(R_D-i+1)(R_D-i)}& 1\leq i\leq l^*,\\
		0& l^*<i\leq R_A;
	\end{cases}\text{~~~~~and~~~~~}
	d^*_j=\begin{cases}
		\displaystyle \frac{1}{R_D-j}& j<l^*,\\
		\displaystyle 1-\sum_{j=0}^{l^*-1}\frac{1}{R_D-j}& j=l^*,\\
		0& l^*<j\leq R_D.
	\end{cases}\]
	The value $l^*$ is given by $l^*=\min(R_A,k^*)$, with $k^*$ identified by: \[H_{R_D}-H_{R_D-k^*} < 1 < H_{R_D}-H_{R_D-k^*-1}.\]
\end{theorem}

\begin{proof}
	The proof is analogous to that of Theorem \ref{theorem:equilibrium}. The differences are as follows: The equation describing Attacker being indifferent between the strategies they use becomes $\pi_A(i+1,d^*)=\pi_A(i,d^*)-1+d_i(R_D-i)$. On Defender's side, the relevant equations become: $R_D-l^*=\pi_D(a^*,l^*)=\pi_D(a^*,0)=a_0R_D$ and $\pi_D(a^*,i)=(1-P_W(a^*,i))(R_D-i)$. Note that the stronger version of Lemma \ref{lemma:suppa=suppd1} is again true: $supp(a^*)=supp(d^*)$, and for the same reason.
\end{proof}

\begin{corollary}
	The sum of expected payoffs is equal to total endowments less the upper bound
	\[\Pi_A+\Pi_D=R_A-E(a^*)+R_D-E(d^*)=R_A+R_D - l^*.\]	
\end{corollary}

\begin{proof}
	From Theorem \ref{uneq:equilibrium}, the expected earnings for Attacker are $R_A$, while for Defender, they are $R_D-l^*$. The sum of these is $R_A+R_D - l^*$.
\end{proof}

\begin{corollary}
	The sum of expected conflict investments is equal to the upper bound \[E(a^*)+E(d^*)=l^*.\]
\end{corollary}
\begin{proof}
	See the proof of Corollary \ref{corollary:sum-conflict-investments}.
\end{proof}

\begin{theorem}
	As Defender's endowment goes to infinity while Attacker's endowment is fixed, the following hold:
	\begin{itemize}
		\item $\lim_{R_D\to \infty} d^*_{R_A}=1.$
		\item $\lim_{R_D\to \infty} a^*_{0}=1.$
		\item $\lim_{R_D\to \infty} P(W)=0.$	
	\end{itemize}
\end{theorem}
\begin{proof}
	Directly from Theorem $\ref{uneq:equilibrium}$, taking limits.
\end{proof}

\begin{theorem}\label{theorem:outside-equilibrium}
	With cost $C\geq 2$ and benefit $B>C$ (and corresponding profit of $Q=B-C$), the A-D game with endowments of $R_A$ and $R_D$ has a single, mixed-strategy Nash equilibrium $(a^*,d^*)$, determined in the following manner.
	
	For Defender, consider first the following sequences:\\
	\noindent Case $1: R_D\geq R_A.$
	\[
	d_j=\begin{cases}
		\displaystyle \frac{1}{R_D+Q-j} & j<R_A-C,\vspace{5pt}\\
		\displaystyle \frac{1+Q}{R_D-R_A+B} & j=R_A-C,\vspace{5pt}\\
		\displaystyle \frac{1}{R_D+Q-j} & R_A-C<j<R_D-C,\vspace{5pt}\\
		\displaystyle \frac{1}{R_D-j} & R_D-C<j\leq R_D
	\end{cases}\]
	Case $2: R_D>R_A.$
	\[
	d_j=\begin{cases}
		\displaystyle \frac{1}{R_D+Q-j} & j\leq R_D-C,\vspace{5pt}\\
		\displaystyle \frac{1}{R_D-j} & R_D-C <j<R_A-C,\vspace{5pt}\\
		\displaystyle \frac{1+Q}{R_D-R_A+C} & j=R_A-C,\vspace{5pt}\\
		\displaystyle \frac{1}{R_D-j} & R_A-C<j\leq R_A
	\end{cases}
	\]
	
	For both cases, let $k^*$ be the value where the cumulative probability reaches or exceeds $1$, i.e. $k^*$ is the smallest value such that $\displaystyle \sum_{j=1}^{l^*}d_j\geq 1$. Then, we set $l^*=\min(k^*, R_A, R_D)$. Defender's equilibrium strategy is then given by:
	
	\[
	d^*_j=\begin{cases}
		\displaystyle d_j & 0\leq j<l^*,\vspace{5pt}\\
		\displaystyle 1-\sum_{j=0}^{l^*-1}d_j& j=l^*,\\
		0& l^*<j\leq R_D.
	\end{cases}
	\]
	
	Given the value of the upper limit $l^*$, we start with the following sequences for generating Attacker's equilibrium strategy:
	
	\noindent Case $I: R_D-l^*\geq C.$
	\[
	a_i=\begin{cases}
		\displaystyle \frac{R_D+Q-l^*}{R_D+Q} & i=0,\vspace{5pt}\\
		\displaystyle \frac{R_D+Q-l^*}{(R_D+Q-i)(R_D+Q-i+1)} & 0<i\leq l^*
	\end{cases}\]
	Case $II: R_D-l^*< C.$
	\[
	a_i=\begin{cases}
		\displaystyle \frac{R_D-l^*}{R_D+Q} & i=0,\vspace{5pt}\\
		\displaystyle \frac{R_D-l^*}{(R_D+Q-i)(R_D+Q-i+1)} & 0<i\leq R_D-C,\vspace{5pt}\\		
		\displaystyle \frac{(R_D-l^*)(1+Q)}{B(C-1)} & i=R_D-C+1,\vspace{5pt}\\		
		\displaystyle \frac{R_D-l^*}{(R_D-i)(R_D-i+1)} & R_D-C+1<i\leq l^*
	\end{cases}
	\]	
	
	Attacker's equilibrium strategy is given by:
	\[
	a^*_i=\begin{cases}
		\displaystyle a_i & 0\leq i<l^*,\vspace{5pt}\\
		\displaystyle 	1-\sum_{i=0}^{l^*-1}a_i& i=l^*,\\
		0& l^*<i\leq R_A.
	\end{cases}
	\]
	
\end{theorem}
\begin{proof}
	Despite the length of the involved formulas, the proof again follows the logic of Theorem \ref{theorem:equilibrium}, and we only need to highlight the differences.
	
	Attacker's expected payoffs $\pi_A(i,d^*)$ depend on whether the remaining resources for either side allow for paying the cost of the outside option $C$. 
	In Case $1$, with $R_A\leq R_D$, the indifference conditions become:
	\begin{align*}
		\pi_A(i+1,d^*)= &\pi_A(i,d^*)-1+d_i(R_D+Q-i)& \text{ for } i<R_D-C, \text{ excepting } i=R_A-C,\\
		\pi_A(i+1,d^*)= &\pi_A(i,d^*)-1-Q+d_i(R_D+Q-i)& \text{ for } i=R_A-C, \text{ and}\\
		\pi_A(i+1,d^*)= &\pi_A(i,d^*)-1+d_i(R_D-i)& \text{ for } i>R_D-C.
	\end{align*}
	In the case of $i=R_A-C$, Attacker is unable to pay for the outside option, and thus loses the  payoff of $Q$ when increasing their contribution from $i$ to $i+1$.
	
	For Case $2$, i.e., when $R_A> R_D$, these conditions become:
	\begin{align*}
		\pi_A(i+1,d^*)&=\pi_A(i,d^*)-1+d_i(R_D+Q-i)& \text{ for } i\leq R_D-C\\
		\pi_A(i+1,d^*)&=\pi_A(i,d^*)-1+d_i(R_D-i)& \text{ for } i>R_D-C\text{, excepting } i=R_A-C.\\
		\pi_A(i+1,d^*)&=\pi_A(i,d^*)-1-Q+d_i(R_D-i)& \text{ for } i=R_A-C.
	\end{align*}
	
	These equations determine a sequence $d_0,d_1,\dotso$. Clearly, the smaller of $R_A$ and $R_D$ put a cap on the distribution; otherwise, the upper bound $l^*$ is determined by the value where the sum of the sequence would exceed $1$, and $d_{l^*}$ will accumulate the remaining probability weight. Hence, $l^*=\min(k^*, R_A, R_D)$, with $k^*$ referring to the cross-over point. 
	
	Turning now to Attacker's equilibrium strategy, in Case $I$, we have $R_D-l^*\geq C$, i.e., Defender can always take the outside option in equilibrium. The equations relevant in determining Attacker's mix become $R_D+Q-l^*=\pi_D(a^*,l^*)=\pi_D(a^*,0)=a_0(R_D+Q)$ and $\pi_D(a^*,i)=(1-P_W(a^*,i))(R_D+Q-i)$ for all $i$. 
	
	In Case $II$, we have $R_D-l^*<C$, i.e., Defender is unable to pay for the outside option when using the upper limit. Attacker's mix is determined by the fact that $R_D-l^*=\pi_D(a^*,l^*)=\pi_D(a^*,0)=a_0(R_D+Q)$ , as well as by:
	\[
	\pi_D(a^*,i)=\begin{cases}
		(1-P_W(a^*,i))(R_D+Q-i)& \text{ for } i\leq R_D-C\\
		(1-P_W(a^*,i))(R_D-i)& \text{ for } i> R_D-C
	\end{cases}
	\]
	At the cross-over between using strategies $R_D-C$ and $R_D-C+1$, Defender's payoff is influenced by both the fact that they need not pay $C$ anymore, as well as not gaining the profits $Q$.
	
	Our definition of $l^*$ guarantees that Attacker's equilibrium mix has the proper support, coinciding with the support of Defender, so Lemma $4$ holds also for the case of outside options.
	
\end{proof}

\end{document}